
\newif\iflanl
\openin 1 lanlmac
\ifeof 1 \lanlfalse \else \lanltrue \fi
\closein 1
\iflanl
    \input lanlmac
\else
    \message{[lanlmac not found - use harvmac instead}
    \input harvmac
\fi
\newif\ifhypertex
\ifx\hyperdef\UnDeFiNeD
    \hypertexfalse
    \message{[HYPERTEX MODE OFF}
    \def\hyperref#1#2#3#4{#4}
    \def\hyperdef#1#2#3#4{#4}
    \def\hypernoname{}
    \def\e@tf@ur#1{}
    \def\hth/#1#2#3#4#5#6#7{{\tt hep-th/#1#2#3#4#5#6#7}}
    \def\CERN{\address{CERN, Geneva, Switzerland}}
\else
    \hypertextrue
    \message{[HYPERTEX MODE ON}
  \def\hth/#1#2#3#4#5#6#7{
  {\tt hep-th/#1#2#3#4#5#6#7}}
\def\CERN{\address{
Theory Division, CERN, Geneva, Switzerland}}
\fi
\newif\ifdraft

\noblackbox
\catcode`\@=11
\newif\iffrontpage
\ifx\answ\bigans
\def\titleft{\titsm}
\magnification=1200\baselineskip=14pt plus 2pt minus 1pt
%
\advance\hoffset by-0.075truein
\advance\voffset by1.truecm
\hsize=6.15truein\vsize=600.truept\hsbody=\hsize\hstitle=\hsize
\else\let\lr=L
\def\titleft{\titla}
\magnification=1000\baselineskip=14pt plus 2pt minus 1pt
%
\hoffset=-0.75truein\voffset=-.0truein
\vsize=6.5truein
\hstitle=8.truein\hsbody=4.75truein
\fullhsize=10truein\hsize=\hsbody
\fi
\parskip=4pt plus 15pt minus 1pt
%
\newif\iffigureexists
\newif\ifepsfloaded
\def\epsfcheck{
\ifdraft
\input epsf\epsfloadedtrue
\else
  \openin 1 epsf
  \ifeof 1 \epsfloadedfalse \else \epsfloadedtrue \fi
  \closein 1
  \ifepsfloaded
    \input epsf
  \else
\immediate\write20{NO EPSF FILE --- FIGURES WILL BE IGNORED}
  \fi
\fi
\def\epsfcheck{}}
\def\checkex#1{
\ifdraft
\figureexistsfalse\immediate%
\write20{Draftmode: figure #1 not included}
\else\relax
    \ifepsfloaded \openin 1 #1
	\ifeof 1
           \figureexistsfalse
  \immediate\write20{FIGURE FILE #1 NOT FOUND}
	\else \figureexiststrue
	\fi \closein 1
    \else \figureexistsfalse
    \fi
\fi}
\def\missbox#1#2{$\vcenter{\hrule
\hbox{\vrule height#1\kern1.truein
\raise.5truein\hbox{#2} \kern1.truein \vrule} \hrule}$}
\def\lfig#1{
\let\labelflag=#1%
\def\numb@rone{#1}%
\ifx\labelflag\UnDeFiNeD%
{\xdef#1{\the\figno}%
\writedef{#1\leftbracket{\the\figno}}%
\global\advance\figno by1%
}\fi{\hyperref{}{figure}{{\numb@rone}}{Fig.{\numb@rone}}}}
\def\figinsert#1#2#3#4{
\epsfcheck\checkex{#4}%
\def\figsize{#3}%
\let\flag=#1\ifx\flag\UnDeFiNeD
{\xdef#1{\the\figno}%
\writedef{#1\leftbracket{\the\figno}}%
\global\advance\figno by1%
}\fi
\goodbreak\midinsert%
\iffigureexists
\centerline{\epsfysize\figsize\epsfbox{#4}}%
\else%
\vskip.05truein
  \ifepsfloaded
  \ifdraft
  \centerline{\missbox\figsize{Draftmode: #4 not included}}%
  \else
  \centerline{\missbox\figsize{#4 not found}}
  \fi
  \else
  \centerline{\missbox\figsize{epsf.tex not found}}
  \fi
\vskip.05truein
\fi%
{\smallskip%
\leftskip 4pc \rightskip 4pc%
\noindent\ninepoint\sl \baselineskip=11pt%
{\bf{\hyperdef\hypernoname{figure}{{#1}}{Fig.{#1}}}:~}#2%
\smallskip}\bigskip\endinsert%
}
%
\font\bigit=cmti10 scaled \magstep1

\font\titla=cmr10 scaled\magstep3
\font\tenmss=cmss10
\font\absmss=cmss10 scaled\magstep1

\newfam\mssfam
\font\footrm=cmr8  \font\footrms=cmr5
\font\footrmss=cmr5   \font\footi=cmmi8
\font\footis=cmmi5   \font\footiss=cmmi5
\font\footsy=cmsy8   \font\footsys=cmsy5
\font\footsyss=cmsy5   \font\footbf=cmbx8
\font\footmss=cmss8
\def\footfont{\def\rm{\fam0\footrm}
\textfont0=\footrm \scriptfont0=\footrms
\scriptscriptfont0=\footrmss
\textfont1=\footi \scriptfont1=\footis
\scriptscriptfont1=\footiss
\textfont2=\footsy \scriptfont2=\footsys
\scriptscriptfont2=\footsyss
\textfont\itfam=\footi \def\it{\fam\itfam\footi}
\textfont\mssfam=\footmss \def\mss{\fam\mssfam\footmss}
\textfont\bffam=\footbf \def\bf{\fam\bffam\footbf} \rm}
\def\tenpoint{\def\rm{\fam0\tenrm}
\textfont0=\tenrm \scriptfont0=\sevenrm
\scriptscriptfont0=\fiverm
\textfont1=\teni  \scriptfont1=\seveni
\scriptscriptfont1=\fivei
\textfont2=\tensy \scriptfont2=\sevensy
\scriptscriptfont2=\fivesy
\textfont\itfam=\tenit \def\it{\fam\itfam\tenit}
\textfont\mssfam=\tenmss \def\mss{\fam\mssfam\tenmss}
\textfont\bffam=\tenbf \def\bf{\fam\bffam\tenbf} \rm}
\ifx\answ\bigans\def\abstractfont{\tenpoint}\else
\def\abstractfont{\def\rm{\fam0\absrm}
\textfont0=\absrm \scriptfont0=\absrms
\scriptscriptfont0=\absrmss
\textfont1=\absi \scriptfont1=\absis
\scriptscriptfont1=\absiss
\textfont2=\abssy \scriptfont2=\abssys
\scriptscriptfont2=\abssyss
\textfont\itfam=\bigit \def\it{\fam\itfam\bigit}
\textfont\mssfam=\absmss \def\mss{\fam\mssfam\absmss}
\textfont\bffam=\absbf \def\bf{\fam\bffam\absbf}\rm}\fi
%
\def\f@@t{\baselineskip10pt\lineskip0pt\lineskiplimit0pt
\bgroup\aftergroup\@foot\let\next}
\setbox\strutbox=\hbox{\vrule height 8.pt depth 3.5pt width\z@}
\def\vfootnote#1{\insert\footins\bgroup
\baselineskip10pt\footfont
\interlinepenalty=\interfootnotelinepenalty
\floatingpenalty=20000
\splittopskip=\ht\strutbox \boxmaxdepth=\dp\strutbox
\leftskip=24pt \rightskip=\z@skip
\parindent=12pt \parfillskip=0pt plus 1fil
\spaceskip=\z@skip \xspaceskip=\z@skip
\Textindent{$#1$}\footstrut\futurelet\next\fo@t}
\def\Textindent#1{\noindent\llap{#1\enspace}\ignorespaces}
\def\foot{\global\advance\ftno by1%
\attach{\hyperref{}{footnote}{\the\ftno}{\footsymbolgen}}%
\vfootnote{\hyperdef\hypernoname{footnote}{\the\ftno}{\footsymbol}}}%
\def\footnote#1{\global\advance\ftno by1%
\attach{\hyperref{}{footnote}{\the\ftno}{#1}}%
\vfootnote{\hyperdef\hypernoname{footnote}{\the\ftno}{#1}}}%
\newcount\lastf@@t           \lastf@@t=-1
\newcount\footsymbolcount    \footsymbolcount=0
\global\newcount\ftno \global\ftno=0
\def\footsymbolgen{\relax\footsym
\global\lastf@@t=\pageno\footsymbol}
\def\footsym{\ifnum\footsymbolcount<0
\global\footsymbolcount=0\fi
{\iffrontpage \else \advance\lastf@@t by 1 \fi
\ifnum\lastf@@t<\pageno \global\footsymbolcount=0
\else \global\advance\footsymbolcount by 1 \fi }
\ifcase\footsymbolcount
\fd@f\dagger\or \fd@f\diamond\or \fd@f\ddagger\or
\fd@f\natural\or \fd@f\ast\or \fd@f\bullet\or
\fd@f\star\or \fd@f\nabla\else \fd@f\dagger
\global\footsymbolcount=0 \fi }
\def\fd@f#1{\xdef\footsymbol{#1}}
\def\space@ver#1{\let\@sf=\empty \ifmmode #1\else \ifhmode
\edef\@sf{\spacefactor=\the\spacefactor}
\unskip${}#1$\relax\fi\fi}
\def\attach#1{\space@ver{\strut^{\mkern 2mu #1}}\@sf}
%
\newif\ifnref
\def\rrr#1#2{\relax\ifnref\nref#1{#2}\else\ref#1{#2}\fi}
\def\ldf#1#2{\begingroup\obeylines
\gdef#1{\rrr{#1}{#2}}\endgroup\unskip}

\def\doubref#1#2{\refs{{#1},{#2}}}

\nreffalse
\def\refout{\listrefs}
%
\def\eqn#1{\xdef #1{(\noexpand\hyperref{}%
{equation}{\secsym\the\meqno}%
{\secsym\the\meqno})}\eqno(\hyperdef\hypernoname{equation}%
{\secsym\the\meqno}{\secsym\the\meqno})\eqlabeL#1%
\writedef{#1\leftbracket#1}\global\advance\meqno by1}
\def\eqnalign#1{\xdef #1{\noexpand\hyperref{}{equation}%
{\secsym\the\meqno}{(\secsym\the\meqno)}}%
\writedef{#1\leftbracket#1}%
\hyperdef\hypernoname{equation}%
{\secsym\the\meqno}{\e@tf@ur#1}\eqlabeL{#1}%
\global\advance\meqno by1}
\def\eqnalign#1{\xdef #1{(\secsym\the\meqno)}
\writedef{#1\leftbracket#1}%
\global\advance\meqno by1 #1\eqlabeL{#1}}
%

%
\def\chap#1{\newsec{#1}}
\def\chapter#1{\chap{#1}}
\def\sect#1{\subsec{#1}}
\def\section#1{\sect{#1}}
\def\\{\ifnum\lastpenalty=-10000\relax
\else\hfil\penalty-10000\fi\ignorespaces}
\def\note#1{\leavevmode%
\edef\@@marginsf{\spacefactor=\the\spacefactor\relax}%
\ifdraft\strut\vadjust{%
\hbox to0pt{\hskip\hsize%
\ifx\answ\bigans\hskip.1in\else\hskip .1in\fi%
\vbox to0pt{\vskip-\dp
\strutbox\sevenbf\baselineskip=8pt plus 1pt minus 1pt%
\ifx\answ\bigans\hsize=.7in\else\hsize=.35in\fi%
\tolerance=5000 \hbadness=5000%
\leftskip=0pt \rightskip=0pt \everypar={}%
\raggedright\parskip=0pt \parindent=0pt%
\vskip-\ht\strutbox\noindent\strut#1\par%
\vss}\hss}}\fi\@@marginsf\kern-.01cm}
\def\titlepage{%
\frontpagetrue\nopagenumbers\abstractfont%
\hsize=\hstitle\rightline{\vbox{\baselineskip=10pt%
{\abstractfont\pubnum}}}\pageno=0}
\frontpagefalse
\def\pubnum{}
\def\pdate{\number\month/\number\yearltd}
\def\makefootline{\iffrontpage\vskip .27truein
\line{\the\footline}
\vskip -.1truein\leftline{\vbox{\baselineskip=10pt%
{\abstractfont\pdate}}}
\else\vskip.5cm\line{\hss \tenrm $-$ \folio\ $-$ \hss}\fi}
\def\title#1{\vskip .7truecm\titlestyle{\titleft #1}}
\def\titlestyle#1{\par\begingroup \interlinepenalty=9999
\leftskip=0.02\hsize plus 0.23\hsize minus 0.02\hsize
\rightskip=\leftskip \parfillskip=0pt
\hyphenpenalty=9000 \exhyphenpenalty=9000
\tolerance=9999 \pretolerance=9000
\spaceskip=0.333em \xspaceskip=0.5em
\noindent #1\par\endgroup }
\def\autskip{\ifx\answ\bigans\vskip.5truecm\else\vskip.1cm\fi}
\def\author#1{\vskip .7in \centerline{#1}}

\def\address#1{\ifx\answ\bigans\vskip.2truecm
\else\vskip.1cm\fi{\it \centerline{#1}}}
\def\abstract#1{
\vskip .5in\vfil\centerline
{\bf Abstract}\penalty1000
{{\smallskip\ifx\answ\bigans\leftskip 2pc \rightskip 2pc
\else\leftskip 5pc \rightskip 5pc\fi
\noindent\abstractfont \baselineskip=12pt
{#1} \smallskip}}
\penalty-1000}
\def\endpage{\tenpoint\supereject\global\hsize=\hsbody%
\frontpagefalse\footline={\hss\tenrm\folio\hss}}
\def\ack{\goodbreak\vskip2.cm\centerline{{\bf Acknowledgements}}}
\def\bfone{\relax{\rm 1\kern-.35em 1}}
\def\inbar{\vrule height1.5ex width.4pt depth0pt}
\def\IC{\relax\,\hbox{$\inbar\kern-.3em{\mss C}$}}
\def\ID{\relax{\rm I\kern-.18em D}}
\def\IF{\relax{\rm I\kern-.18em F}}
\def\IH{\relax{\rm I\kern-.18em H}}
\def\II{\relax{\rm I\kern-.17em I}}
\def\IN{\relax{\rm I\kern-.18em N}}
\def\IP{\relax{\rm I\kern-.18em P}}
\def\IQ{\relax\,\hbox{$\inbar\kern-.3em{\rm Q}$}}
\def\IR{\relax{\rm I\kern-.18em R}}
\font\cmss=cmss10 \font\cmsss=cmss10 at 7pt
\def\ZZ{\relax\ifmmode\mathchoice
{\hbox{\cmss Z\kern-.4em Z}}{\hbox{\cmss Z\kern-.4em Z}}
{\lower.9pt\hbox{\cmsss Z\kern-.4em Z}}
{\lower1.2pt\hbox{\cmsss Z\kern-.4em Z}}\else{\cmss Z\kern-.4em
Z}\fi}
\def\a{\alpha} \def\b{\beta}

\def\L{\Lambda} 
 
\def\cC{{\cal C}} 
\def\cF{{\cal F}}

\def\cL{{\cal L}}

\def\nup#1({Nucl.\ Phys.\ $\us {B#1}$\ (}
\def\plt#1({Phys.\ Lett.\ $\us  {#1}$\ (}
\def\cmp#1({Comm.\ Math.\ Phys.\ $\us  {#1}$\ (}
\def\prp#1({Phys.\ Rep.\ $\us  {#1}$\ (}
\def\prl#1({Phys.\ Rev.\ Lett.\ $\us  {#1}$\ (}
\def\prv#1({Phys.\ Rev.\ $\us  {#1}$\ (}
\def\mpl#1({Mod.\ Phys.\ Let.\ $\us  {A#1}$\ (}
\def\ijmp#1({Int.\ J.\ Mod.\ Phys.\ $\us{A#1}$\ (}
\def\tit#1|{{\it #1},\ }
%

%

\def\ni{\noindent}
\def\tilde{\widetilde}
\def\bar{\overline}
\def\us#1{\underline{#1}}

\def\Coe#1.#2.{{#1\over #2}}
\def\coeff#1#2{\relax{\textstyle {#1 \over #2}}\displaystyle}
\def\coe#1.#2.{\relax{\textstyle {#1 \over #2}}\displaystyle}

\def\shalf{\relax{\textstyle {1 \over 2}}\displaystyle}

\def\to{\rightarrow}
\def\notin{\hbox{{$\in$}\kern-.51em\hbox{/}}}

\def\del{\partial}

\def\nex#1{$N\!=\!#1$}

\catcode`\@=12
\def\a{a_1}
\def\ai{a_i}
\def\b{a_2}
\def\ad{a_{D;1}}
\def\bd{a_{D;2}}
\def\adi{a_{D;i}}
\def\cF{{\cal F}}
\def\wan#1{W_{\!A_{#1}}}
\def\D{\Delta}
\def\cW{{\cal W}}
\def\rc{r^{{\rm class}}}
\def\cC{{\cal C}}

%
\ldf\SWa{N.\ Seiberg and E.\ Witten, \nup426(1994) 19, \hth/9407087.}
\ldf\SWb{N.\ Seiberg and E.\ Witten, {\it Monopoles, Duality, and
Chiral Symmetry Breaking in N=2 Supersymmetric QCD}, preprint
RU-94-60, IASSNS-94-55, \hth/9408099.}
\ldf\LGrefs{E. Martinec, \plt 217B(1989) 431;
C.\ Vafa and N.P.\ Warner, \plt218B  (1989) 51.}
\ldf\KLTY{A.\ Klemm, W.\ Lerche, S.\ Theisen
and S. Yankielowicz, in preparation.}
\ldf\Arn{See e.g., V.\ Arnold, A.\ Gusein-Zade and A.\ Varchenko,
{\it Singularities of Differentiable Maps I, II}, Birkh\"auser 1985.}
\ldf\theUltimateSpec{P.\ Townsend, \plt202B (1988) 53; C.\ Hull and
P.\ Townsend, {\it Unity of superstring dualities}, preprint
QMW-94-30, \hth/9410167.}
\def\DVV {\rrr\DVV {R.\ Dijkgraaf, E. Verlinde and H. Verlinde,
\nup352(1991) 59.}}
\ldf\specgeo{S.\ Ferrara and A.\ Strominger, in ``Strings 89'', eds.\
R.\ Arnowitt et al. (World Scientific, Singapore, 1989), p.\ 245.}
\ldf\CAF{A.\ Ceresole, R.\ D'Auria and S.\ Ferrara, \plt339B(1994)
71,
\hth/9408036.}
\ldf\NS{N.\ Seiberg, \nup303(1988) 286.}
\ldf\BR{R.\ Brandt and F.\ Neri, \nup161 (1979) 253.}
\ldf\NSbeta{N.\ Seiberg, \plt206B(1988) 75.}
\ldf\SCHIF{
V.\ Novikov, M.\ Schifman, A.\ Vainstein,
M\ Voloshin and V.\ Zakharov,
\nup229(1983) 394;
V.\ Novikov, M.\ Schifman, A.\ Vainstein and V.\ Zakharov,
\nup229(1983) 381, 407;
M.\ Schifman, A.\ Vainstein and V.\ Zakharov,
\plt166B(1986) 329.}
%
\def\pubnum{
\hbox{CERN-TH.7495/94}
\hbox{LMU-TPW 94/16}
\hbox{hep-th/9411048}}
\def\pdate{}
\titlepage
\title
{Simple Singularities and N=2 Supersymmetric
Yang-Mills Theory}
\vskip-1.1cm\autskip
\author{A.\ Klemm, W.\ Lerche, S.\ Yankielowicz\footnote {a}
{On leave of absence from the School of Physics, Raymond and Beverly
Sackler Faculty of Exact Sciences, Tel-Aviv University.}
$\!\!{}^,$\footnote b {Work supported in part by the US-Israel
Binational Science Foundation and GIF - the German-Israeli Foundation
for Scientific Research.}}
\CERN
\vskip .5truecm
\centerline{and}
\vskip-1.7truecm
\author{S.\ Theisen$^b$}
\address{Sektion Physik, University Munich, Germany}
\vskip-1.truecm
\abstract{We present a first step towards generalizing the work of
Seiberg and Witten on \nex2 supersymmetric Yang-Mills theory to
arbitrary gauge groups. Specifically, we propose a particular
sequence of hyperelliptic genus $n\!-\!1$ Riemann surfaces to underly
the quantum moduli space of $SU(n)$ \nex2 supersymmetric gauge
theory. These curves have an obvious generalization to arbitrary
simply laced gauge groups, which involves the A-D-E type simple
singularities. To support our proposal, we argue that the monodromy
in the semiclassical regime is correctly reproduced. We also give
some remarks on a possible relation to string theory.}
\vfil
\vskip 1.cm
\ni CERN-TH.7495/94\hfill\break
\ni November 1994
\endpage
\baselineskip=14pt plus 2pt minus 1pt
\sequentialequations
\chapter{Introduction}

In two beautiful papers \doubref\SWa\SWb, Seiberg and Witten have
investigated \nex2 supersymmetric $SU(2)$ gauge theories and solved
for their exact nonperturbative low energy effective action. For
arbitrary gauge group $G$, such supersymmetric theories are
characterized by having flat directions for the Higgs vacuum
expectation values, along which the gauge group is generically broken
to the Cartan subalgebra. Thus, the effective theories contain
$r\!=\!{\rm rank}(G)$ abelian \nex2 vector supermultiplets, which can
be decomposed into $r$ \nex1 chiral multiplets $A^i$ plus $r$ \nex1
vector multiplets $W^i_\alpha$. The \nex2 supersymmetry implies that
the effective theory depends only on a single holomorphic
prepotential $\cF(A)$. More precisely, the effective lagrangian in
\nex1 superspace is
$$
\cL\ =\ {1\over4\pi}{\rm Im}\,\Big[\, \int \!d^4\theta\,\big(\sum
{\del
\cF(A)\over\del A^i}\bar A^i\big) \,+\, \int \!d^2\theta\,{1\over2}
\big(\sum {\del^2 \cF(A)\over\del A^i\del
A^j}W_\alpha^iW_\alpha^j\big)\Big]\ . \eqn\effL
$$
The holomorphic function $\cF$ determines the quantum moduli space
and, in particular, its metric. This space has singularities at
points or surfaces where additional fields become massless, that is,
where the effective action description breaks down. A crucial insight
is that the electric and magnetic quantum numbers of the fields that
become massless at a given singularity are determined by the
eigenvalue-(+1) eigenvectors of the monodromy matrix associated with
the singularity.

For $G=SU(2)$ considered in \doubref\SWa\SWb, besides the point at
$u=\infty$ there are singularities at $u=\pm\L^2$, where $\L$ is the
dynamically generated scale of the theory, and $u=\shalf \langle
a^2\rangle$, where $a\equiv A|_{\theta=0}$. (On the other hand, $u=0$
is not singular in the exact quantum theory, which means that, in
contrast to the classical theory, no massless non-abelian gauge
bosons arise here). One of these points corresponds to a massless,
purely magnetically charged monopole, and the other to a massless
dyon. The parameter region near $u=\infty$ describes the
semiclassical, perturbative regime, which is governed by the one-loop
beta function \doubref\SCHIF\NSbeta. It gives rise to a non-trivial
monodromy as well (arising from the logarithm in the one-loop beta
function), but there are no massless states associated with it.

The singularity structure and knowledge of the monodromies allow to
completely determine the holomorphic prepotential $\cF$. The
monodromy group is $\Gamma(2)\subset SL(2,\ZZ)$ consisting of all
matrices congruent to $\bfone$ modulo 2. The matrices act on the
vector $(a_D;a)^t$, where $a_D$ is the magnetic dual of $a$, that is,
$a_D\equiv {\del F(a)\over\del a}$. The quantum moduli space, namely
the $u$-plane punctured at $\pm \L^2$ and $\infty$, can thus be
thought of $\IH/\Gamma(2)$, where $\IH$ is the upper half-plane.

The basic idea \SWa\ in solving for the effective theory is to
consider the following family of holomorphic curves parametrized by
$\IH/\Gamma(2)$:
$$
y^2\ =\ (x-\L^2)(x+\L^2)(x-u)\ .
\eqn\curve
$$
These curves represent a double cover of the $x$-plane with branch
points at $0,\pm\L^2$ and $\infty$, and describe a genus one Riemann
surface. That is, the quantum moduli space of the $SU(2)$ super
Yang-Mills theory coincides with the moduli space of a particular
torus; this torus becomes singular when two branch points in \curve\
coincide. The derivatives of the electric and magnetic coordinates
$(a_D;a)^t$ with respect of $u$ are just given by the periods.
Computing the period integrals (related to the two homology cycles)
thus yields, upon integration, the dependence of $a, a_D$ in terms of
$u$, and integrating $a_D$ finally determines the prepotential
$\cF(a)$.

In the present paper, we make a first step towards generalizing the
work of Seiberg and Witten to (pure) super Yang-Mills theory with
$SU(n)$ gauge group (we also will hint at how it might work for
arbitrary simply laced groups). More precisely, we will propose what
we think the appropriate curves are, and give circumstantial evidence
to the fact that our choice is correct. We will present a more
detailed analysis of the monodromies and period integrals in a
follow-up paper \KLTY.

\chapter{Semiclassical Regime}

To be specific, we will consider mainly the gauge group
$G=SU(3)$, but from our setup it will be clear that
all of our arguments immediately generalize to arbitrary
$SU(n)$. We will denote the gauge invariant order parameters
(Casimirs) by
$$
u\ =\ \shalf {\rm Tr}\,\langle\,\phi^2\,\rangle\ ,\qquad\qquad\ \
v\ =\ \coeff13 {\rm Tr}\,\langle\,\phi^3\,\rangle\ ,
\eqn\uvdef
$$
where we can always take the scalar superfield component to be
$\phi={\rm diag}(\a,\b-\a,-\b)$, such that, classically,
$u={\a}^2+{\b}^2-\a\b$, $v=\a\b(\a-\b)$. The residual global $\ZZ_6$
symmetry act as $u\to e^{2\pi i/3}u$, $v\to -v$. For generic
eigenvalues of $\phi$, the $SU(3)$ gauge symmetry is broken to
$U(1)\times U(1)$, whereas if any two eigenvalues are equal, the
unbroken symmetry is $SU(2)\times U(1)$. These classical symmetry
properties are encoded in the following, gauge and globally $\ZZ_6$
invariant discriminant:
$$
\D_0\ =\ 4 u^3 - 27 v^2\ =\ (\a+\b)^2(2\a-\b)^2(\a-2\b)^2\
\ .\eqn\clasD
$$
The lines $\D_0=0$ in $(u,v)$ space correspond to unbroken
$SU(2)\times U(1)$, and have a cusp singularity at the origin, where
the $SU(3)$ symmetry is restored. As we will see, in the full quantum
theory the cusp is smoothed out, $\D_0\to\D_\L = 4 u^3 - 27 v^2 +
O(\L^3)$, which, in particular, prohibits a phase with massless
non-abelian gluons.

It is straightforward to compute the prepotential $\cF$
in the perturbative regime, with the result
$$
\cF_{{\rm class}}\ =\ {i\over 4\pi}
\sum_{i<j}^3(e_i-e_j)^2\log[(e_i-e_j)^2/\L^2] \ .
\eqn\clasF
$$
Here, $e_i$ denote the roots of the equation
$$
\wan2(x,u,v)\ \equiv\ x^3 - u\,x - v\ =\ 0\ ,
\eqn\Wsing
$$
whose bifurcation set is given by $\D_0$ in \clasD. Whenever two
roots
coincide, the discriminant vanishes. In terms of
the variables $\a,\b$ we have:
$$
\eqalign{
e_1-e_2\ &=\ (\a+\b)\cr
e_1-e_3\ &=\ (2\a-\b)\cr
e_2-e_3\ &=\ (\a-2\b)\cr
}\eqn\eiarel
$$
and thus, accordingly
$$
\wan2(x,\a,\b)\ = (x-\a)(x-(-\b))(x-(\b-\a))\ .
\eqn\Wab
$$
The Casimirs $u,v$ are gauge invariant and, in particular,
invariant under the Weyl group $\cW$ of $SU(3)$. This group is
generated
by any two of the reflections
$$
\eqalign{
r_1:\ \  (\a,\b)\ &\rightarrow\ (\b-\a,\b)\cr
r_2:\ \  (\a,\b)\ &\rightarrow\ (\a,\a-\b)\cr
r_3:\ \  (\a,\b)\ &\rightarrow\ (-\b,-\a)\ .\cr
}\eqn\weyldef
$$
Due to the multi-valuedness of the inverse map
$(u,v)\to(\a,\b)$,closed paths in $(u,v)$ space will in general not
close in $(\a,\b)$ space, but will in general close only up to Weyl
transformations. Such a monodromy will be non-trivial if a given path
encircles a singularity in $(u,v)$ space --- in our case, the
singularities will be at ``infinity'' and along the lines where the
discriminant vanishes.

It is indeed well-known \Arn\ that the monodromy group of the simple
singularity of type $A_2$ \Wsing\ is given by the Weyl group of
$SU(3)$, and acts as Galois group on the $e_i$ (and analogously for
$\wan{n-1}$ related to $SU(n)$). This will be the starting point for
our generalization.

What we are interested in, of course, is not just the monodromy
acting on $(\a,\b)$, but the monodromy acting on $(\ad,\bd;\a,\b)^t$,
where
$$
a_{D,i}\ \equiv\ \cF_i\ =\ {\del\over\del a_i}\,\cF(\a,\b)\ .
\eqn\aD
$$
Performing the Weyl reflection $r_1$ on $(\a,\b)^t$, we easily
find
$$
\left(\matrix{\cF_1\cr\cF_2\cr}\right)\ \rightarrow\
\pmatrix{-1 & 0\cr 1 & 1 \cr}\left(\matrix{\cF_1\cr\cF_2\cr}\right)+
N\, \pmatrix{2 & -1\cr -1 &
-1\cr}\left(\matrix{\a\cr\b\cr}\right)\ .
\eqn\ronead
$$
The winding number $N$ that arises from the logarithms is not
determined by the finite, ``classical'' Weyl transformation acting on
the $a_i$, but depends on the chosen path in $(u,v)$ space. We note
that for large $Z\equiv 4 u^3 - 27 v^2$, which corresponds to the
semiclassical limit, the prepotential behaves like
$$
\cF_{{\rm class}}\sim u\,\log \big[{Z\over \L^6}\big]\ ,
\eqn\Fc
$$
from which we can read off the winding number for any given path in
the semiclassical regime. We find that by choosing appropriate paths,
one can have $N$ jump by even integers, and that the minimal winding
number is \nex1. (An example\foot {We need to assume here that this
path really exists quantum mechanically.} for such a closed loop is
given by $(u(a_i(t)),v(a_i(t))$ for $t=0,\dots,1$, where
$a_1(t)=e^{i\pi t}a_1 +\shalf(1-e^{i\pi t})a_2$, $a_2(t)=a_2$.)
Therefore, the matrix representation of $r_1$ acting on
$(\ad,\bd;\a,\b)^t$ is:
$$
r_1\ =\ \pmatrix{ -1 & 0 & 2 & -1 \cr 1 & 1 & -1 & -1 \cr 0 & 0 & -1
& 1
   \cr 0 & 0 & 0 & 1 \cr }\ \equiv\ \rc_1\, T^{-1}\ ,
\eqn\ronedef
$$
where $\rc_1$ is the ``classical'' Weyl reflection (given by the
block diagonal part of $r_1$), and $T$ the ``quantum monodromy''
$$
T\ =\ \pmatrix{\textstyle\bfone & C\cr 0&\bfone\cr},\ \ \qquad{\rm
where}\ \ C\ =\ \pmatrix{\textstyle2 & -1\cr-1&2\cr}
\eqn\quantmon
$$
is the Cartan matrix of
$SU(3)$. The other Weyl reflections are given analogously by
$r_i=\rc_i\,T^{-1}$. The $r_i$ are related to each other by
conjugation,
and, in particular, rotate into each other via the Coxeter
element, $\rc_{{\rm cox}}=\rc_1\rc_2$.

\chapter{The Curves for $SU(n)$}

Our aim is now to find a sequence of curves $\cC$ that reproduces the
quantum moduli space of supersymmetric $SU(n)$ Yang-Mills theories.
Since we do not know how to derive these curves from first
principles, we will make a proposal for the curves that is consistent
with various requirements, and subsequently verify that at least the
monodromies at ``infinity'' reproduce the above matrices $r_i$. This
will be our only non-trivial consistency check for the time being. To
really show that the choice of curves is correct physicswise,
requires in addition to check the various other monodromies, which
correspond to the condensation of monopoles and dyons. A detailed
discussion of these matters will be presented elsewhere \KLTY.

Let us now list the requirements that we impose on the curves $\cC$.
First, we seek surfaces with $2n$ periods (corresponding to
$(\adi;\ai)$), whose period matrices are positive definite. It will
be pointed out below that this condition can be satisfied by choosing
genus $n-1$ Riemann surfaces, in direct generalization of \SWa.
Secondly, we require that for $\L\to0$ the classical situation is
recovered. That is, the discriminant of $\cC$ should have, for
$\L=0$, a factor of $\D_0$ given in \clasD. This means that for
$\L=0$ the curves should have the form $y^m=\cC(x)\equiv\wan{n-1}\!\!
\times\!(\dots)$ for some $m$. This then also implies that the
monodromy groups will have something to do with the Weyl groups of
$SU(n)$, and this is what we want as well. Thirdly, the curves must
behave properly under the cyclic global transformations acting on the
Casimirs $\{c_2,c_3,\dots\}\equiv\{u,v,\dots\}$; in other words,
there should be a natural dependence on Casimirs, for all groups.
Finally, from \SWb\ we know that $\L$ should appear in $\cC(x)$ with
a power that corresponds to the charge violation of the one-instanton
process.

Taking these requirements together suggests the surfaces for
$SU(n)$ Yang-Mills theory to be the following, genus $g\!=\!n\!-\!1$
hyperelliptic curves:
$$
y^2\ =\ \cC_n(x)\ \equiv\ \Big(\wan{n-1}(x,c_i)\Big)^2 - \L^{2n}\ ,
\eqn\theCurve
$$
where
$$
\wan{n-1}(x,c_i)\ =\ x^n - \sum_{i=2}^n c_i\,x^{n-i}.
\eqn\LGpots
$$
are the $A$-type simple singularities related to $SU(n)$. Morally
speaking, the square of $W$ reflects having both electric and
magnetic degrees of freedom. Since there is a general relationship
\Arn\ between Arnold's simple A-D-E singularities, perturbations by
Casimirs, monodromy and Weyl groups, we conjecture that \theCurve\
describes surfaces for the other simply laced gauge groups $G$ as
well, by simply replacing $\wan{n-1}(x,c_i)$ by the corresponding D-
or E-type singularity, and $\L^{2n}$ by $\L^{2h}$, where $h$ is the
corresponding Coxeter number.

Note in passing that even though $\cC_2(x)$ does not have the form as
one of the curves given in \doubref\SWa\SWb, it is equivalent to the
$\Gamma_0(4)$ modular curve given in \SWb, since the modular
invariants\foot{Obtained by transforming to the Weierstrass normal
form.} coincide: $j(u)={1\over27\L^8} {(3\L^4-4u^2)^3
\over\L^4-u^2}$. Note also that the points $u^2=\L^4$ and $u=\infty$
are exchanged for the two curves in \doubref\SWa\SWb, i.e., the
parameters of the $\Gamma(2)$ and $\Gamma_0(4)$ modular curves are
related as follows:
$$
u_{(4)}\ =\ {u_{(2)}\over \sqrt{{u_{(2)}}^2-\L^4}}\L^2\ .
\eqn\urel
$$

\ni Returning to $SU(n)$, it is useful to write
$$
\eqalign{
\cC_n(x)\ &=\
\Big(\wan{n-1}(x,c_i)+\L^n\Big)\Big(\wan{n-1}(x,c_i)-\L^n\Big)\cr
&=\ \prod_{i=1}^n(x-e_i^+)(x-e_i^-)\ .\cr}
\eqn\factcurve
$$
Critical surfaces occur whenever two roots of $\cC(x)$ coincide, that
is, whenever the discriminant $\D_\L=\prod_{i<j}(e_i^\pm-e_j^\pm)^2$
vanishes. Physicswise we expect when this happens, monopoles or dyons
condense whose quantum numbers are determined by the corresponding
monodromy matrices. For example, for $G=SU(3)$ the quantum
discriminant is:
$$
\D_\L\ =\ \L^{18}\D_\L^+\D_\L^-\ ,  \ \ \
\ \ \D^\pm_\L\ =\ 4 u^3-27(v\pm\L^3)^2.
\eqn\thefullD
$$

By construction, the hyperelliptic curves \theCurve\ are represented
by branched covers over the $x$-plane. More precisely, we have $n$
$\ZZ_2$ cuts, each linking a pair of roots $e^+_i$ and $e^-_i$,
$i=1,\dots,n$. As an example, we present the picture for $G=SU(3)$ in
\lfig\figone. In the classical theory, where $\L\to0$, the branch
lines shrink to $n$ doubly degenerate points: $e^-_i\to e^+_i\equiv
e_i$. These points, given for $SU(3)$ in eq.\ \eiarel, correspond to
the weights of the $n$-dimensional fundamental representation (the
picture represents a deformed projection of the weights onto the
unique Coxeter eigenspace with $\ZZ_n$ action). This means that the
branched $x$-plane transforms naturally under the finite
``classical'' Weyl group that permutes the points. This finite Weyl
group is all there is in the classical theory, and is just the usual
monodromy group of the $A_{n-1}$ singularity alluded to earlier. In
the quantum theory, where the degenerate dots are resolved into
branch lines, there are in addition possibilities for ``quantum
monodromy'', which involves braiding of the cuts.
\figinsert\figone{Branched $x$-plane with cuts
linking pairs of roots of $\cC_3=0$. We depicted our choice of basis
for the homology cycles. Condensation of monopoles or dyons occurs
when two branch points approach each other. The monodromy of the
corresponding vanishing cycle then determines the electric and
magnetic quantum numbers.
}{2.5in}{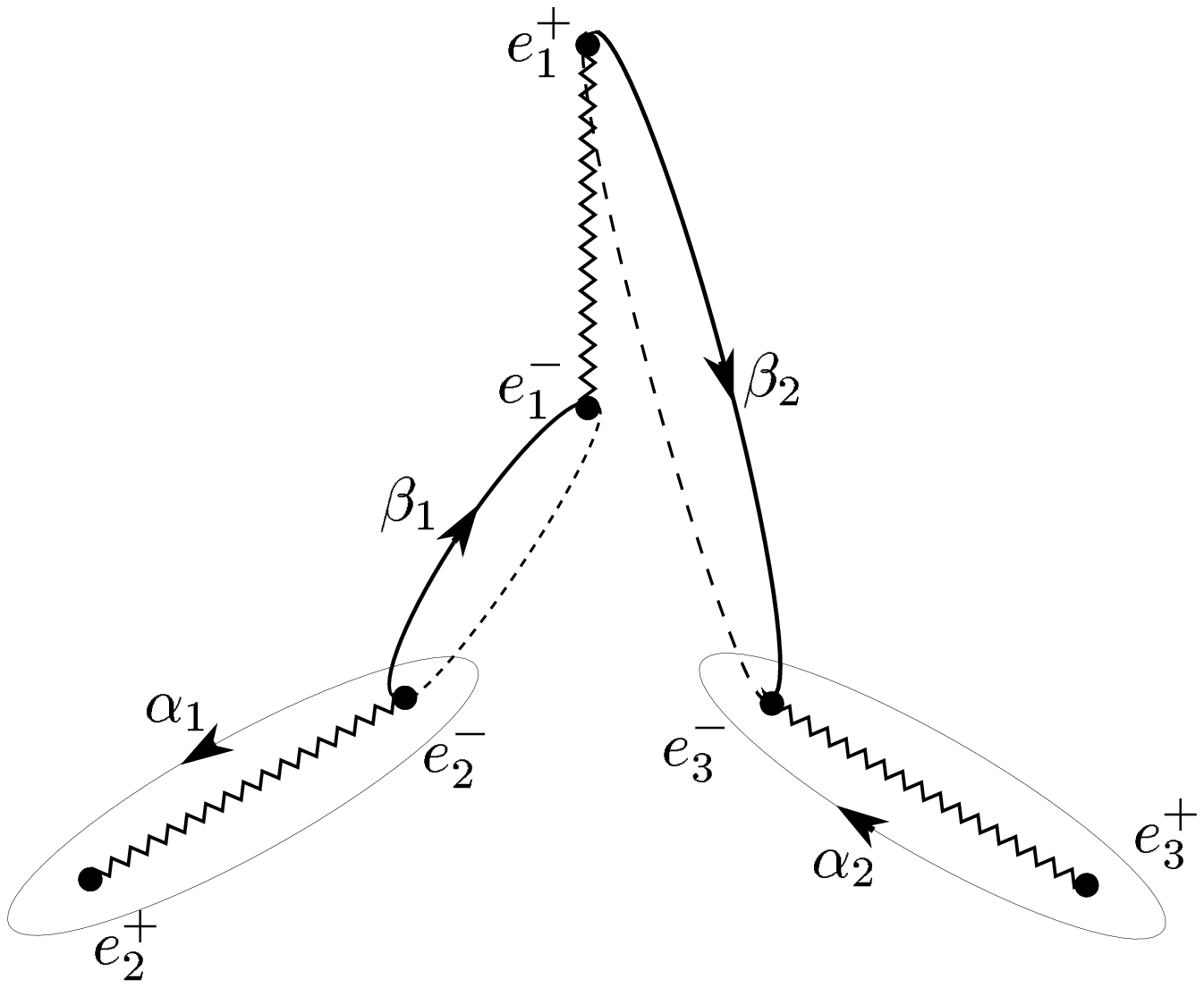}

\chapter{Periods and Monodromies for G=SU(3)}

What we are interested in is the monodromy at infinity, which happens
to be an ``unstable'' situation in that more than two points collide
simultaneously (at $x=\infty$). This would in principle require to
find an appropriate compactification of the moduli space to make the
degeneration stable.\foot{For $G=SU(2)$, one can do this by just
fixing three points on the $x$-plane. Doing this has the effect that
$u=\pm\L^2$ and $u=\infty$ get exchanged, precisely according to the
reparametrization \urel. Thus, what one might call monodromy at
infinity gets exchanged with what one might call monodromy at
$u=\L^2$. The monodromy matrices, however, are essentially the same
(up to conjugation and inversion), so the difference does not seem to
matter.} However, instead of trying to resolve this subtle problem,
we will rather employ a trick to get at the monodromy at infinity in
a more direct way.

Specifically, we will use the fact the the monodromy factors into a
classical and a quantum part, just as in \ronedef\
for $SU(3)$. From the above it is quite clear that the classical part
of the monodromy is obtained by simply permuting the branches in
the $x$-plane. This is easily implemented by choosing appropriate
paths
just like the one above eq.\ \ronedef.

The quantum part is associated with the logarithm in \Fc. The crucial
observation is that we can mimic the effect of looping around in the
$Z$-plane (where $Z\equiv4u^3-27v^2$), by formally rotating
$$
{\cal T}:\ \ \L^{6}\ \to\ e^{2\pi i t}\L^{6}\ ,\qquad\ \ t=0,\dots,1\
,
\eqn\loop
$$
along a small cycle around the origin. Such a rotation of a
singularity $\tilde W=W(x,c_i=0)+\epsilon$, $\epsilon=e^{2\pi i
t}$, is indeed well-known in the mathematical literature \Arn, where
it is, ironically, called ``classical monodromy''. For the A-D-E
simple singularities it corresponds to the Coxeter element of the
Weyl group, whereas in the present context, it gives the quantum
monodromy in the semiclassical regime.

Now what $\cal T$ does on the $x$-plane is to transform the $e_i^+$
and $e^-_i$ into each other --- this is obvious from \factcurve.
Therefore, the quantum monodromy is given by the product of all the
monodromy matrices associated with the vanishing cycles around the
branch cuts in the $x$-plane. So what needs to be done is to
determine the precise form of these matrices.

Clearly, the monodromy matrices must reflect the action of braiding
and permuting the cuts on the vector $(\adi;\ai)^t$. This action is
expressed in terms of the action on the homology cycles via
$$
\adi\ =\ \oint_{\beta_i}\lambda\ ,\ \ \ \ \ \
\ai\ =\ \oint_{\alpha_i}\lambda\ ,
\eqn\aints
$$
where $\alpha_i,\beta_j$ is some symplectic homology basis with
$\langle\alpha_i,\beta_j\rangle=-\langle\beta_j,\alpha_i \rangle=
\delta_{ij}$, $\langle\alpha_i, \alpha_j\rangle=\langle\beta_i,
\beta_j\rangle=0$, $i,j=1,\ldots,g$. From the theory of Riemann
surfaces it is clear that the monodromy group must be contained in
$Sp(2g,\ZZ)=Sp(2n\!-\!2,\ZZ)$. For $G=SU(3)$, we have depicted our
choice of homology basis in \lfig\figone.

In \aints, $\lambda$ denotes a suitably chosen meromorphic
differential. The holomorphic differentials on a genus $g=n-1$
hyperelliptic curve are given by $\omega_i= x^{i-1}{dx\over y}$,
$i=1,\ldots,g$. They give rise to the period matrices as
$A_{ij}=\int_{\alpha_j}\!\omega_i$ and
$B_{ij}=\int_{\beta_j}\!\omega_i$, which are related to $(\adi,\ai)$
as follows:
$$
A_{ij}={\partial a^i(u)\over \partial u_j},
\quad B_{ij}={\partial a_{D}^i(u)\over \partial u_j}\ ,
\eqn\Ident
$$
This represents a non-trivial integrability condition, and this is
what determines $\lambda$ in \aints. Specifically, we can have that
$\omega_i={\partial \lambda\over \partial u_i}$ if we (formally)
choose, for example, the differential as follows:
$$
\lambda
\ =\ -dx\,{\rm log}[-W_{A_{n-1}}-
\sqrt{(W_{A_{n-1}})^2-\Lambda^{2 n}}]\ .
\eqn\lam
$$
Note that due to the identification \Ident,
the second Riemann bilinear relation, ${\rm Im} (A^{-1}B)>0$,
ensures the positivity of the metric:
$$
(ds)^2={\rm Im} {\partial^2 {\cal F}\over \partial a^i\partial a^j}
d a^i d \bar a^j={\rm Im } \sum_{i=1}^g d a_{D\, i} d \bar a^i\ ,
\eqn\metric
$$
similar as for genus one \SWa.

The action of braiding the branch points on the homology can be
obtained elementarily by tracing the deformations of the cycles
induced by the movement of the branch points. Somewhat easier and
more elegant is however to use the Picard-Lefshetz formula \Arn.
Denoting by $\nu_{ij}$ the vanishing cycle that vanishes as one moves
the $i$'th branch point along a specified path to the $j$'th point,
the action on any cycle $\gamma$ of the counter-clockwise braiding of
the two points along that path is given by
$$
S_{\nu_{ij}}\gamma=\gamma+\langle\gamma,\nu_{ij}\rangle
\nu_{ij}\ ,
\eqn\picardle
$$
where $\langle\gamma,\nu_{ij}\rangle$ denotes the intersection
of the two cycles. (For the A-D-E simple singularities,
this formula coincides with the well-known formula
for Weyl reflections.)

Specifically, the effect of braiding the cut between $e_1^+$ and
$e_1^-$ on our homology basis, defined in \lfig\figone, comes out to
be as follows:
$$
B_{e_1^+,e_1^-}\ =\
\pmatrix{ 1 & 0 & -1 & 1 \cr 0 & 1 & 1 & -1 \cr 0 & 0 & 1 & 0 \cr 0
   & 0 & 0 & 1 \cr  }
$$
There is no additional sign since the forms $\omega_i$ are invariant
for this particular braid. Similarly,
$$
\eqalign{
B_{e_2^+,e_2^-}\ &=\
\pmatrix{ 1 & 0 & -1 & 0 \cr 0 & 1 & 0 & 0 \cr 0 & 0 & 1 & 0 \cr 0 &
  0 & 0 & 1 \cr  }\cr
B_{e_3^+,e_3^-}\ &=\
\pmatrix{ 1 & 0 & 0 & 0 \cr 0 & 1 & 0 & -1 \cr 0 & 0 & 1 & 0 \cr 0 &
  0 & 0 & 1 \cr  }
}$$
According to what we said above, the quantum monodromy is then given
by the product of these matrices, and it indeed coincides with
\quantmon:
$$
B_{e_1^+,e_1^-}\,B_{e_2^+,e_2^-}\,B_{e_3^+,e_3^-}\ \equiv\ T^{-1}\ .
\eqn\goodnews
$$
The classical monodromy depends on the particular path, and is
trivially given by the corresponding classical Weyl group element.
For example, for the path given above eq.\ \ronedef, the branches $1$
and $2$ in \lfig\figone\ are exchanged, and the action of this braid
on the homology is given by $\rc_1$. Hence, the full monodromy is
$r_1=\rc_1T^{-1}$, in accordance with eq.\ \ronedef. We believe that
this is a non-trivial verification of our proposal for the
hyperelliptic curves.

\chapter{Remarks about a relation to string theory}

The A-D-E singularities play also a well-known r\^ole in string
theory \LGrefs. They give rise to exactly solvable d=2 \nex2
supersymmetric Landau-Ginzburg models, whose tensor products can be
used to represent Calabi-Yau string compactifications. We can indeed
relate our curves as well to LG models, by simply going to homogenous
coordinates. The LG superpotentials are of the form
$W_{LG}=y^2+x^{2n}+(\L z)^{2n}+ c_2 x^{2n-2} z^2 +\dots$, where $c_i$
are now dimensionless moduli, and $\L$ is an irrelevant, non-zero
number. These potentials describe tensor products of two \nex2
minimal models of type $A_{2n-1}$.

As is well-known \DVV, such \nex2 theories, when viewed as
topological field theories, are characterized by prepotentials
$\cF_{LG}(\tau(a))$, where $\tau(a(c))$ are the flat coordinates
corresponding to the LG moduli $c$. The point is that the computation
of $\cF_{LG}$ is more or less the same as the computation that leads
to $\cF$ in \effL, and therefore these two prepotentials are very
closely related. Therefore, if we consider type IIB string
compactification with \nex2 space-time supersymmetry in $d=4$, on a
superconformal background that contains one of the above LG models as
a tensor product piece of it, the string effective action
\doubref\NS\specgeo\ contains a piece that is very similar to the the
\nex2 Yang-Mills effective action \effL. It might be possible that,
upon decoupling the gravitational sector (``rigid special
geometry''\doubref\SWa\CAF) and appropriately freezing the various
other fields, the effective actions do coincide. It is indeed
well-known that the abelian gauge group in the RR-sector of a type II
string compactification never enlarges to a non-abelian group, and
this may be thought as a reflection of what happens for quantum \nex2
Yang-Mills theory.

Does this potentially mean that a low-energy observer cannot
distinguish between this subsector of the type IIB string
compactification and the effective Yang-Mills theory ? The answer
might be related to the conjecture \theUltimateSpec\ about the
equivalence of string theory with its effective field theory, when
all solitons of the effective theory are taken into account. The
above would also imply that the same, specific complex functional
dependence of a given $\cF$ on the moduli could be attributed either
to world-sheet instanton effects (that is, to properties of 2d CFT),
or, equally well, to space-time non-perturbative effects (ie., to 4d
super Yang Mills theory), and such a relation would seem to be quite
non-trivial.

\ni We believe that these matters urgently deserve further study.

\ack

We understand that Luis \ Alvarez-Gaum\'e has independently come to
conclusions similar to the ones presented in this paper, and thank
him, as well as Sergio Ferrara, Dennis Nemeschansky and
Shing-Tung Yau, for useful comments.

\refout
\end